\documentclass{obspm}
\usepackage{graphicx}
\begin{document}

\TitreGlobal{Hunt for Molecules}

\title{Hunt for cold H$_2$ molecules}
\author{Combes, F.}\address{Observatoire de Paris, LERMA, 
61 Av. de l'Observatoire, F-75014, Paris}
\runningtitle{Hunt for cold H$_2$ molecules}
\setcounter{page}{23}
\index{Combes, F.}
%
\begin{abstract} 
The bulk of the molecular component in galaxies is made
of cold H$_2$, which is not observed
directly, but which abundance is derived from indirect 
tracers such as CO emission. The CO to H$_2$
conversion ratio remains uncertain, and may vary by large
factors in special environments with different 
excitation or metallicity. Recent cold gas discoveries
(through $\gamma$-rays or cold dust emission)
are reviewed and the most promising tracers
in the future are discussed, such as the primordial 
molecules HD and LiH, or the pure rotational lines
of excited H$_2^*$.  
\end{abstract}
\maketitle
%
\section{Introduction}

The H$_2$ molecule has no electric dipole because of its symmetry,
and therefore cannot emit line radiation in the radio domain,  
the only ones that could be excited at the low temperature of the
interstellar medium (10-15K). The first rotational line
comes from the $J=2$ level, through quadrupole radiation,
and is 512K above the ground state, S(0) at 28$\mu$m in wavelength.
The second rotational line of the para H$_2$ is S(2) at 12$\mu$m, and 
the two first lines of the ortho H$_2$ are S(1) and S(3) at
17$\mu$m and 9$\mu$m respectively.
ISO observations of PDR have revealed a large number
of H$_2$ pure rotational lines (Timmermann et al 1996), and higher
resolution observations have confirmed an
unexpectedly large amounts of warm H$_2$ gas in PDR (Allers et al 2005).

The only other way to detect cold H$_2$ directly is through the UV lines 
in absorption, unfortunately this concerns only the untypical very diffuse
regions, since higher H$_2$ column densities are associated to high
extinction, that prevent to detect the background sources. This
was confirmed by the FUSE survey towards
galactic sources (Shull et al 2004). The average molecular fraction of the
diffuse ISM is of the order of 10\%, and the temperature is warm ($\sim$ 100K).
Even in "translucent line of sights", there is only evidence of
diffuse clouds (Rachford et al 2002).

We first review our current knowledge of the CO to H$_2$ conversion ratio,
which is still the main method to trace cold H$_2$ molecules in galaxies.
After presenting possible new tracers,
 we then describe recent discoveries of cold gas, locally in our
Galaxy, and in external galaxies or clusters.
 We conclude with future experiments, which could bring
new insights.

\section{CO and other tracers}

\subsection{Uncertainties in the CO to H$_2$ conversion ratio}

 Although the determination of this factor has been the goal
of multiple studies, it remains largely uncertain; it can
vary by a factor 10 or more with metallicity, and also
with excitation and local density. When studied for
individual clouds, from virial mass arguments, the
conversion factor varies by a factor 10 across the mass
spectrum of observed clouds.
For extreme starbursts, it is usual to adopt a conversion
ratio 5 times lower than the average, and the value to
adopt for high-z or primordial galaxies is unknown. 

Since the factor could be large in the outer parts of galaxies,
at low metallicity, it is tempting to wonder how much 
the dark H$_2$ gas could account for the galaxy dark matter,
in particular the dark baryons (i.e. 90\% of them).
It is indeed known from microlensing experiments that
most of them cannot be in compact objects, and must be
in gas, either hot or cold.

The rotation curves here serve as an upper limit on the
conversion ratio.
In spiral and dwarf galaxies, rotation curves are remarkably fit
through HI-scaling by a factor 7-10 (Hoekstra et al 2001).
So inside the HI disk, i.e. inside the last point of directly measured
rotation curve, it is possible to have only
10-20\% of the dark baryons in galaxies.
The remaining part must reside in cosmic filaments.

\subsection{Other possible tracers}

To reveal cold H$_2$ molecules, we could think of
their hyperfine structure (or ultrafine),
coming from the interaction between nuclear spin, and 
the magnetic field generated by rotation.
The ground state has 3 levels (F = I+J= 0, 1 and 2)
and the corresponding lines are in the kilometric range:
 F=1-0 at 546.4 kHz  ($\lambda$ =0.5km) and
F=2-1 at 54.8 kHz ($\lambda$= 5.5km).
The line intensities are extremely weak
 (cf Combes \& Pfenniger 1997), the radiation is stopped by the
atmosphere, and their detection would require
to cover by a loose grid several kilometers on the Moon!
This also assumes that there is a significant amount
of molecules in the ortho state, even at low temperatures,
i.e. the molecules have no time to reach complete
equilibrium after their formation.

At 3K, the pressure of H$_2$ clumps is 100 times the saturated
vapor pressure (Combes \& Pfenniger 1997). However, since the
latent heat is 110K/H$_2$, there is no time to form much snow
in a Hubble time. 
But the conditions for dimerisation  might be reached
in some very dense clumps, and 
continuum emission is expected from dimers, through the
dipole induced by collisions (Schaefer 1994, 1999).
It is difficult to estimate the amount of H$_2$
at such high densities, and the amount of dimers formed.

The HD molecule has a weak dipole moment
$\mu$= 5.8 10$^{-4}$ Debye (Trefler \& Gush 1968). The
first rotational level is
130 K above ground state, and the first $J=1-0$ line is
at $\lambda$ = 112$\mu$m (Wright et al 1999, 
Polehampton et al 2002). The emission could come
only from excited regions, and is quite weak,
given that HD/H$_2$ is of the order 10$^{-5}$. 
HD is therefore not a powerful tracer.
The LiH molecule could be a better tracer, since its
dipole moment is much stronger
$\mu$= 5.9 Debye (Lawrence et al 1963), and the first
transition $J=1-0$ is only 21 K above ground state, 
at $\lambda$ = 0.67$\mu$m (450GHz). This line
cannot be detected from the ground, because of H$_2$O absorption.
Assuming an abundance of LiH/H$_2$ $\sim 10^{-10}$, the line
becomes optically thick $\tau \sim$ 1 for
 N(LiH) = 10$^{12}$ cm$^{-2}$ or N(H$_2$) = 10$^{22}$ cm$^{-2}$.

The ion H$_2^+$ (searched for by Encrenaz \& Falgarone,
as soon as 1971!) has an expected
abundance of 10$^{-11}$-10$^{-10}$. It has an
hyperfine structure, but not in the N=0 state, only N=1, I=1, S=1/2.
The energy of the upper level is 110K, among the 5 lines expected,
the strongest is at 1343 MHz.
If H$_2^+$ forms through cosmic ray ionization of H$_2$, 
it disappears through reaction with H$_2$ to form the molecular ion
 H$_3^+$, which has a key role in ion-molecule
interstellar chemistry (e.g. Geballe 2000).
The H$_3^+$ ion has now been detected in the interstellar medium
(Geballe \& Oka 1996, McCall et al 1999),
through absorption lines in the infra-red, via a vibrationnally
excited line (since only the asymmetric vibration $\nu_2$
can induce a dipole moment, in this ion which has no
permanent dipole). The deuterated ions are more favorable
tracers, since they have a weak dipole, 
H$_2$D$^+$ (372GHz emission, Stark et al 1999), or D$_2$H$^+$ 
(692GHz emission, Vastel et al 2004).

Given the weakness of all these tracers, it might be
interesting to keep tracking the CO molecule, since
traces of C and O are still expected in nearly primordial gas.
A residual 10$^{-3}$ solar abundance is observed in Ly$\alpha$ 
forest clouds, due to the first stars.
If the H$_2$ gas remains clumpy enough, and far
from photo-dissociating radiation, it is possible
to find CO molecules, assuming that some
heating sources exist to keep the brightness 
temperature above the background (e.g. Braine \& Herpin 2004).
Alternatively, the molecules should be searched
in absorption, although with a low
surface filling factor of less than 1\%.

\section{Hints of dark gas}

\subsection{$\gamma-$rays}

\begin{figure}[h]
 \centering
 \includegraphics[width=8cm]{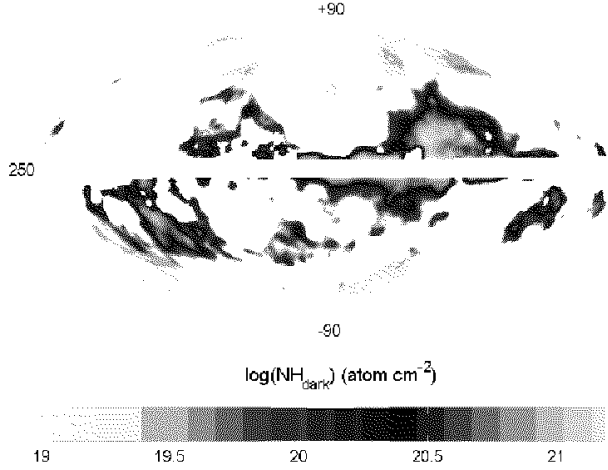}
 \caption{Map of the local dark gas, in Galactic coordinates centered on l=70$^\circ$, 
as traced from $\gamma$-ray intensity, reddening E(B-V) and 
94 GHz emission, from Grenier et al (2005)}
    \label{gamma}
   \end{figure}

Another way to trace cold gas in the interstellar medium
is the $\gamma$-rays resulting from the interaction of
cosmic-rays (CR) with all nucleons (through the creation of $\pi^0$).
The emission of E $>$ 100 Mev $\gamma$-rays is the product of
total nucleon density and CR density. The CR are 
re-accelerated by supernovae and are the products of star formation,
their density decreases radially in the Galaxy. It is possible
to calibrate their density in the solar neighbourhood, through 
mapping of all ISM components (HI, CO, H$\alpha$). Until recently,
due to the low spatial resolution, it was possible to attribute
part of the $\gamma$-rays to point sources (pulsars), but 
converging observations now, including extinction (B-V) and
cold dust emission in the millimeter range, reveals that
$\gamma$-rays are tracing dark gas in the local interstellar
medium (Grenier et al 2005). This interesting discovery
shows that the dark gas is distributed like an envelope
around the CO-traced molecular gas, cf
Figure \ref{gamma}.

\subsection{High Velocity Clouds}

Some of the HVC falling towards the Galactic plane appear
compatible with an intergalactic origin, with 
low metallicity (Wakker et al 1999) and high deuterium abundance
(Sembach et al 2004).
 Although H$_2$ molecules have been detected through UV absorption
by FUSE in HVC, up to now no CO emission nor cold dust emission
had been detected. For the first time, Miville-Desch\^enes et al (2005) report
dust emission through the comparison of SST (Spitzer Space Telescope) infrared maps with
the 21cm HI line emission from the GBT (Green Bank Telescope). The dust
is cooler ($\sim$ 10K) than the galactic dust in average, compatible with a 
large distance from the Galaxy. The column density derived is 5 times that
corresponding to the HI, suggesting the existence of dense gas clumps associated
to the apparent diffuse HI gas. The clumps would contain dense molecular
hydrogen, and would constitute most of the gas accreted by the Galaxy. 

\subsection{Cooling flows}

\begin{figure}[h]
 \centering
 \includegraphics[width=9cm]{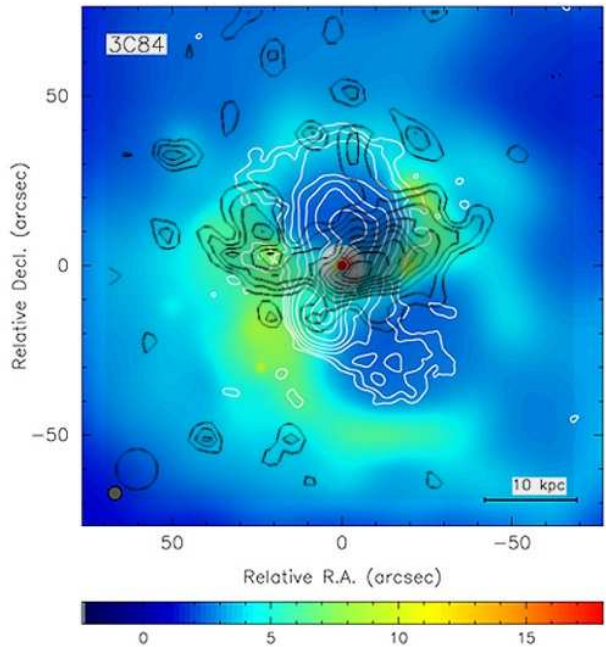}
 \caption{Map of the center of Perseus cluster: the false color-scale is
the X-ray emission from Chandra (Fabian et al 2003), the white contours are the
radio emission (Pedlar et al 1990), showing the radio lobes of the central AGN,
and the black contours are the CO(2-1) emission, observed with 
the IRAM-30m (from Salom\'e et al 2005).}
    \label{cool}
   \end{figure}

Cold gas has also recently been discovered in the center of rich galaxy
clusters, where it was long searched for, as the end product of cooling flows.
 It has been realized that the flux of cooling gas is in reality about
10 times lower than expected in the case of spherical inflow without feedback.
 The cold inflow fuels an AGN in the central cD galaxy, and the radio lobes
feedback then moderates the cooling, creating bubbles and shocks
(Figure \ref{cool}). The cold gas has still the dynamics and velocity
of the cluster, and is not yet settled in the central
galaxy potential well (see Salom\'e et al 2005, and Salom\'e, this
Proceedings).

\section{Perspectives}

The dark H$_2$ clumps are not compact enough to be
gravitational lenses for LMC stars, however they could
act as gaseous lenses (Draine 1998) or
produce scintillation of optical light, through refraction
of the background stellar emission (Moniez 2003).
In the strong diffractive regime, it might be possible to distinguish high contrast fluctuations
of short time-scales from foreground effects of light propagation through the atmosphere. A
project is undertaken to monitor extra-galactic stars every $\sim$ 10 seconds, to discover dark gas
of 10$^{19}$ molecules/cm$^2$ per 10 000 km transverse distance
(OSER: Optical Scintillation by Extraterrestrial Refractors, 
Moniez 2005).
 
The pure rotational lines of the H$_2$ molecule might be the best tracer
afterall.  ISO observations have shown that they are ubiquitous.
In outer parts of galaxies they were observed 
surprisingly high by Valentijn \& Van der Werf (1999).
In the edge-on galaxy NGC891, the molecular column
density was derived to be 
N(H$_2$)= 10$^{23}$ cm$^{-2}$, with an excitation temperature of
T = 80-90 K, i.e. 5-15 \, N(HI), enough to explain the rotation curve.

 ISO observations of lines of sights through the Galaxy encountering no star formation
regions, but only cold diffuse gas, have shown a rich variety of H$_2$ rotational
lines. The excitation of these lines in the ground vibrational states cannot be due 
principally to the UV excitation, but has to come from non-thermal excitation
(Falgarone et al 2005a). This suggests that within the 
cold neutral medium, there always exists a few percent of warm gas, due to the intermittent 
dissipation of MHD turbulence, and this warm gas can be used as a tracer of the cold gas,
independent of metallicity. 
  The detection of this warm gas, at large scale in the Galaxy,
and in a survey of nearby galaxies, in particular at large distance
from the center, is the main goal of the project H$_2$Ex, a space explorer
mission with a dedicated 2m-class telescope, and a high resolution 
spectro-imager in the mid-IR (Falgarone et al 2005b).

  The rotational lines of H$_2$ are the main coolant of the primordial gas,
and its collapse to form the first generation of stars in the universe could
be traced by their emission, however orders of magnitude gains in sensitivity
are required for future instruments (Mizusawa et al 2005).



\begin{thebibliography}{}
\bibitem{}Allers, K. N., Jaffe, D. T., Lacy et al. : 2005, ApJ 630, 368
\bibitem{}Braine J., Herpin F.: 2004 Nature 432, 369
\bibitem{}Combes F., Pfenniger D.: 1997, A\&A  327, 453
\bibitem{}Draine B.T: 1998 ApJ 509, L41
\bibitem{}Encrenaz P.J., Falgarone E.: 1971 ApL 8, 187
\bibitem{}Fabian A.C., Sanders J.S., Allen S.W. et al: 2003, MNRAS 344, L43
\bibitem{}Falgarone, E., Verstraete, L., Pineau Des For\^ets, G., Hily-Blant, P.: 2005a, A\&A 433, 997
\bibitem{}Falgarone, E., Boulanger, F., Combes, F. et al.: 2005b, EAS Publications Series, 
Vol 14, 2005, pp.57-66
\bibitem{}Geballe T.R.: 2000, Phil. Trans. Royal Soc. A, 358, 2503-2514
\bibitem{}Geballe T.R., Oka T.: 1996, Nature 384, 334
\bibitem{}Grenier, I.A., Casandjian, J.-M., Terrier, R.: 2005, Science 307, 1292
\bibitem{}Hoekstra, H., van Albada, T. S., Sancisi, R.: 2001, MNRAS  323, 453
\bibitem{}Lawrence T.R., Anderson C.H., Ramsey N.F.: 1963, Phys Rev 130, 1865
\bibitem{}McCall B.J., Geballe T.R., Hinkle K.H., Oka T.: 1999, ApJ 522, 338
\bibitem{}Miville-Desch\^enes, M.-A., Boulanger, F., Reach, W. T., Noriega-Crespo, A.: 2005, ApJ 631, L57
\bibitem{}Mizusawa, H., Nishi, R., Omukai, K.: 2005, PASJ 56, 487
\bibitem{}Moniez M.: 2003 A\&A 412, 105
\bibitem{}Moniez M.: 2005  in "Mass Profiles and Shapes of Cosmological Structures", 
           21st IAP colloquium, EdP Sciences, in press
\bibitem{}Pedlar A., Ghataure H.S., Davies R.D. et al.:1990, MNRAS 246, 477
\bibitem{}Polehampton, E. T., Baluteau, J.-P., Ceccarelli, C., Swinyard, B. M., Caux, E.: 2002 A\&A 388, L44
\bibitem{}Rachford, B. L., Snow, T. P., Tumlinson, J. et al: 2002, ApJ 577, 221
\bibitem{}Salom\'e P., Combes F., Edge A. et al: 2005, A\&A Letter, in prep.
\bibitem{}Schaefer J.: 1994, A\&A 284, 1015
\bibitem{}Schaefer J.: 1999, in H$_2$ in Space, Eds.: F. Combes, G. Pineau des For\^ets, Cambridge Univ. Press
\bibitem{}Sembach, K. R., Wakker, B. P., Tripp, T. M. et al: 2004, ApJS 150, 387
\bibitem{}Shull, J. M., Anderson, K., Tumlinson, J. et al.: 2004, BAAS 204, 6118
\bibitem{}Stark, R., van der Tak, F.S., van Dishoeck, E.F.: 1999, ApJ 521, L67
\bibitem{}Timmermann, R., Bertoldi, F., Wright, C. M. et al: 1996, A\&A 315, L281
\bibitem{}Trefler M., Gush H.P.: 1968, Phys Rev Let 20, 708
\bibitem{}Valentijn, E. A., van der Werf, P. P.: 1999, ApJ  522, L29
\bibitem{}Vastel, C., Phillips, T. G., Yoshida, H.: 2004, ApJ 606, L127
\bibitem{}Wakker, B. P., Howk, J. C., Savage, B. D. et al: 1999, Nature  402, 388
\bibitem{}Wright, C.M., van Dishoeck, E.F., Cox, P. et al.: 1999, ApJ 515, L29
\end{thebibliography}
\end{document}